\documentclass[twocolumn,showpacs,preprintnumbers,amsmath,amssymb,floatfix,prd,superscriptaddress]{revtex4}

\usepackage{graphicx}
\usepackage{dcolumn}
\usepackage{bm}
\usepackage{lineno}
\usepackage{setspace}
\usepackage{revsymb}
\usepackage{url}
\usepackage{epsfig}

\newcommand{\met}{\mbox{$E_{T}\!\!\!\!\!\!\!/\,\,\,\,$}}
\newcommand{\cl}{\mbox{C.\ \!\!L.\ \!\!}}
\newcommand{\cll}{\mbox{C.\ \!\!L}}

\begin{document}


%
%
\title{\large \bf \boldmath Search for New Dielectron Resonances and Randall-Sundrum \\Gravitons at the Collider Detector at Fermilab}
\affiliation{Institute of Physics, Academia Sinica, Taipei, Taiwan 11529, Republic of China} 
\affiliation{Argonne National Laboratory, Argonne, Illinois 60439, USA} 
\affiliation{University of Athens, 157 71 Athens, Greece} 
\affiliation{Institut de Fisica d'Altes Energies, ICREA, Universitat Autonoma de Barcelona, E-08193, Bellaterra (Barcelona), Spain} 
\affiliation{Baylor University, Waco, Texas 76798, USA} 
\affiliation{Istituto Nazionale di Fisica Nucleare Bologna, $^{aa}$University of Bologna, I-40127 Bologna, Italy} 
\affiliation{University of California, Davis, Davis, California 95616, USA} 
\affiliation{University of California, Los Angeles, Los Angeles, California 90024, USA} 
\affiliation{Instituto de Fisica de Cantabria, CSIC-University of Cantabria, 39005 Santander, Spain} 
\affiliation{Carnegie Mellon University, Pittsburgh, Pennsylvania 15213, USA} 
\affiliation{Enrico Fermi Institute, University of Chicago, Chicago, Illinois 60637, USA}
\affiliation{Comenius University, 842 48 Bratislava, Slovakia; Institute of Experimental Physics, 040 01 Kosice, Slovakia} 
\affiliation{Joint Institute for Nuclear Research, RU-141980 Dubna, Russia} 
\affiliation{Duke University, Durham, North Carolina 27708, USA} 
\affiliation{Fermi National Accelerator Laboratory, Batavia, Illinois 60510, USA} 
\affiliation{University of Florida, Gainesville, Florida 32611, USA} 
\affiliation{Laboratori Nazionali di Frascati, Istituto Nazionale di Fisica Nucleare, I-00044 Frascati, Italy} 
\affiliation{University of Geneva, CH-1211 Geneva 4, Switzerland} 
\affiliation{Glasgow University, Glasgow G12 8QQ, United Kingdom} 
\affiliation{Harvard University, Cambridge, Massachusetts 02138, USA} 
\affiliation{Division of High Energy Physics, Department of Physics, University of Helsinki and Helsinki Institute of Physics, FIN-00014, Helsinki, Finland} 
\affiliation{University of Illinois, Urbana, Illinois 61801, USA} 
\affiliation{The Johns Hopkins University, Baltimore, Maryland 21218, USA} 
\affiliation{Institut f\"{u}r Experimentelle Kernphysik, Karlsruhe Institute of Technology, D-76131 Karlsruhe, Germany} 
\affiliation{Center for High Energy Physics: Kyungpook National University, Daegu 702-701, Korea; Seoul National University, Seoul 151-742, Korea; Sungkyunkwan University, Suwon 440-746, Korea; Korea Institute of Science and Technology Information, Daejeon 305-806, Korea; Chonnam National University, Gwangju 500-757, Korea; Chonbuk National University, Jeonju 561-756, Korea} 
\affiliation{Ernest Orlando Lawrence Berkeley National Laboratory, Berkeley, California 94720, USA} 
\affiliation{University of Liverpool, Liverpool L69 7ZE, United Kingdom} 
\affiliation{University College London, London WC1E 6BT, United Kingdom} 
\affiliation{Centro de Investigaciones Energeticas Medioambientales y Tecnologicas, E-28040 Madrid, Spain} 
\affiliation{Massachusetts Institute of Technology, Cambridge, Massachusetts 02139, USA} 
\affiliation{Institute of Particle Physics: McGill University, Montr\'{e}al, Qu\'{e}bec, Canada H3A~2T8; Simon Fraser University, Burnaby, British Columbia, Canada V5A~1S6; University of Toronto, Toronto, Ontario, Canada M5S~1A7; and TRIUMF, Vancouver, British Columbia, Canada V6T~2A3} 
\affiliation{University of Michigan, Ann Arbor, Michigan 48109, USA} 
\affiliation{Michigan State University, East Lansing, Michigan 48824, USA}
\affiliation{Institution for Theoretical and Experimental Physics, ITEP, Moscow 117259, Russia}
\affiliation{University of New Mexico, Albuquerque, New Mexico 87131, USA} 
\affiliation{Northwestern University, Evanston, Illinois 60208, USA} 
\affiliation{The Ohio State University, Columbus, Ohio 43210, USA} 
\affiliation{Okayama University, Okayama 700-8530, Japan} 
\affiliation{Osaka City University, Osaka 588, Japan} 
\affiliation{University of Oxford, Oxford OX1 3RH, United Kingdom} 
\affiliation{Istituto Nazionale di Fisica Nucleare, Sezione di Padova-Trento, $^{bb}$University of Padova, I-35131 Padova, Italy} 
\affiliation{LPNHE, Universite Pierre et Marie Curie/IN2P3-CNRS, UMR7585, Paris, F-75252 France} 
\affiliation{University of Pennsylvania, Philadelphia, Pennsylvania 19104, USA}
\affiliation{Istituto Nazionale di Fisica Nucleare Pisa, $^{cc}$University of Pisa, $^{dd}$University of Siena and $^{ee}$Scuola Normale Superiore, I-56127 Pisa, Italy} 
\affiliation{University of Pittsburgh, Pittsburgh, Pennsylvania 15260, USA} 
\affiliation{Purdue University, West Lafayette, Indiana 47907, USA} 
\affiliation{University of Rochester, Rochester, New York 14627, USA} 
\affiliation{The Rockefeller University, New York, New York 10065, USA} 
\affiliation{Istituto Nazionale di Fisica Nucleare, Sezione di Roma 1, $^{ff}$Sapienza Universit\`{a} di Roma, I-00185 Roma, Italy} 

\affiliation{Rutgers University, Piscataway, New Jersey 08855, USA} 
\affiliation{Texas A\&M University, College Station, Texas 77843, USA} 
\affiliation{Istituto Nazionale di Fisica Nucleare Trieste/Udine, I-34100 Trieste, $^{gg}$University of Trieste/Udine, I-33100 Udine, Italy} 
\affiliation{University of Tsukuba, Tsukuba, Ibaraki 305, Japan} 
\affiliation{Tufts University, Medford, Massachusetts 02155, USA} 
\affiliation{University of Virginia, Charlottesville, VA  22906, USA}
\affiliation{Waseda University, Tokyo 169, Japan} 
\affiliation{Wayne State University, Detroit, Michigan 48201, USA} 
\affiliation{University of Wisconsin, Madison, Wisconsin 53706, USA} 
\affiliation{Yale University, New Haven, Connecticut 06520, USA} 
\author{T.~Aaltonen}
\affiliation{Division of High Energy Physics, Department of Physics, University of Helsinki and Helsinki Institute of Physics, FIN-00014, Helsinki, Finland}
\author{B.~\'{A}lvarez~Gonz\'{a}lez$^w$}
\affiliation{Instituto de Fisica de Cantabria, CSIC-University of Cantabria, 39005 Santander, Spain}
\author{S.~Amerio}
\affiliation{Istituto Nazionale di Fisica Nucleare, Sezione di Padova-Trento, $^{bb}$University of Padova, I-35131 Padova, Italy} 

\author{D.~Amidei}
\affiliation{University of Michigan, Ann Arbor, Michigan 48109, USA}
\author{A.~Anastassov}
\affiliation{Northwestern University, Evanston, Illinois 60208, USA}
\author{A.~Annovi}
\affiliation{Laboratori Nazionali di Frascati, Istituto Nazionale di Fisica Nucleare, I-00044 Frascati, Italy}
\author{J.~Antos}
\affiliation{Comenius University, 842 48 Bratislava, Slovakia; Institute of Experimental Physics, 040 01 Kosice, Slovakia}
\author{G.~Apollinari}
\affiliation{Fermi National Accelerator Laboratory, Batavia, Illinois 60510, USA}
\author{J.A.~Appel}
\affiliation{Fermi National Accelerator Laboratory, Batavia, Illinois 60510, USA}
\author{A.~Apresyan}
\affiliation{Purdue University, West Lafayette, Indiana 47907, USA}
% ...
\collaboration{CDF Collaboration\footnote{With visitors from $^a$University of Massachusetts Amherst, Amherst, Massachusetts 01003,
$^b$Istituto Nazionale di Fisica Nucleare, Sezione di Cagliari, 09042 Monserrato (Cagliari), Italy,
$^c$University of California Irvine, Irvine, CA  92697, 
$^d$University of California Santa Barbara, Santa Barbara, CA 93106
$^e$University of California Santa Cruz, Santa Cruz, CA  95064,
$^f$CERN,CH-1211 Geneva, Switzerland,
$^g$Cornell University, Ithaca, NY  14853, 
$^h$University of Cyprus, Nicosia CY-1678, Cyprus, 
$^i$University College Dublin, Dublin 4, Ireland,
$^j$University of Fukui, Fukui City, Fukui Prefecture, Japan 910-0017,
$^k$Universidad Iberoamericana, Mexico D.F., Mexico,
$^l$Iowa State University, Ames, IA  50011,
$^m$University of Iowa, Iowa City, IA  52242,
$^n$Kinki University, Higashi-Osaka City, Japan 577-8502,
$^o$Kansas State University, Manhattan, KS 66506,
$^p$University of Manchester, Manchester M13 9PL, England,
$^q$Queen Mary, University of London, London, E1 4NS, England,
$^r$University of Melbourne, Victoria 3010, Australia,
$^s$Muons, Inc., Batavia, IL 60510,
$^t$Nagasaki Institute of Applied Science, Nagasaki, Japan, 
$^u$National Research Nuclear University, Moscow, Russia,
$^v$University of Notre Dame, Notre Dame, IN 46556,
$^w$Universidad de Oviedo, E-33007 Oviedo, Spain, 
$^x$Texas Tech University, Lubbock, TX  79609, 
$^y$Universidad Tecnica Federico Santa Maria, 110v Valparaiso, Chile,
$^z$Yarmouk University, Irbid 211-63, Jordan,
$^{hh}$On leave from J.~Stefan Institute, Ljubljana, Slovenia, 
}}
\noaffiliation


\date{June 6, 2011}

\begin{abstract}

A search for new dielectron mass resonances using data recorded by the CDF II detector
and corresponding to an integrated luminosity of 5.7 fb$^{-1}$ is presented.  No significant excess over 
the expected standard model prediction is observed.
In this dataset, an event with the highest dielectron mass ever observed (960~GeV/$c^2$) was recorded.
The results are interpreted in the Randall-Sundrum (RS) model.  Combined with the 5.4 fb$^{-1}$
diphoton analysis, the RS-graviton lower mass limit for the coupling $k/\overline{M}_{Pl}=0.1$ 
is 1058~GeV/$c^2$, making it the strongest limit to date.

\end{abstract}

\maketitle
Although extremely successful, the standard model (SM) of particles and fields is not sufficient for answering 
many open physics questions including the origin of the dark matter, the incorporation of gravity, 
and the weak-scale/Planck-scale hierarchy.  Many extensions of the SM have been proposed, most of which predict 
the existence of new particles.  The most direct way to discover a new unstable particle is through 
a mass resonance of its decay products.  In this Letter we present a search for high-mass dielectron resonances.  

At hadron colliders, the search for high-mass dilepton resonances has the advantage of very low hadronic backgrounds 
and well-understood electroweak backgrounds.  
The search for dielectron resonances is additionally motivated by the event excess at a dielectron mass 
$M_{ee}=240$~GeV/$c^2$ reported by a CDF analysis \cite{cdf_ee} performed with an integrated luminosity of 2.5 fb$^{-1}$.
Moreover, a recent CDF search \cite{cdf_gg} for Randall-Sundrum (RS) 
gravitons decaying to diphotons will benefit significantly from the addition of the dielectron decay channel.  
The RS model \cite{RS} solves the hierarchy problem by introducing an extra compact dimension accessible 
only to gravity.  The phenomenology of this model leads to a small number of distinct Kaluza-Klein states (spin-2 gravitons $G$) that couple to SM particles and can be individually detected as resonances of pairs of jets, leptons, 
photons or gauge bosons.  Other possible sources of dielectron resonances include the production of heavy spin-1 (e.g., $Z'$ \cite{zprime}) and
spin-0 particles (e.g., Higgs boson \cite{higgs} and supersymmetric particles \cite{susy}).  

In this Letter we first present a model-independent 
search for dielectron resonances and we subsequently use the results to exclude part of the RS-graviton parameter space.  
The dielectron search is combined with the recent CDF diphoton RS-graviton search to set the strongest limits to date.  Searches for 
dielectron resonances with RS-graviton interpretation have been previously published by the CDF \cite{cdf_ee} and D0 \cite{d0} collaborations.

CDF II \cite{cdfDetector} is a multi-purpose cylindrical detector with projective-tower 
calorimeter geometry and excellent lepton identification capability.  It operates at Fermilab's Tevatron collider where protons and antiprotons collide with a center-of-mass energy of 1.96 TeV.
In CDF's coordinate system the positive $\vec{z}$-axis is defined by the proton beam direction and the positive $\vec{y}$-axis by the vertical upward 
direction.  The detector is approximately symmetric in the $\eta$ and $\phi$ directions, where the pseudorapidity 
$\eta$ is defined as $\eta = -\ln[\tan (\theta/2)]$, $\theta$ is the polar angle with respect to $\vec{z}$, and $\phi$ is the azimuthal angle.

The momentum $p$ of charged particles is measured with a tracking system composed of an eight-layer silicon
strip detector and a 96-layer drift chamber;
both are located inside a solenoid providing a magnetic field of $1.4$~T aligned 
along the beam axis.  
The tracking efficiency is nearly 100\% in the 
central region ($|\eta|<1$) and decreases in the forward region ($1<|\eta|<2.8$).  
Electromagnetic and hadronic calorimeters surround the solenoid and measure the energies of collision products up to $|\eta|=3.6$. 
Gas Cherenkov counters measure the average number of inelastic $p\bar{p}$ collisions 
per bunch crossing and thereby determine the beam luminosity.  A pipelined three-level trigger 
system \cite{trigger} that combines hardware and software is used for filtering 
the collision data.

We perform an analysis of CDF dielectron data collected 
with a high-transverse momentum ($p_T\equiv p\sin\theta >18$~GeV/$c$) central-electron trigger and corresponding
to an integrated luminosity of 5.7 fb$^{-1}$.  To ensure a uniform trigger response,
we require the central electron to have $p_T>20$~GeV/$c$.
A supplementary trigger, more efficient especially 
for electrons with $p_T>70$~GeV/$c$, is also used.  The second electron can pass
either through the central or forward detector region and it is required to have $p_T>5$~GeV/$c$.  
The tracking system provides the 
direction of the electrons, whereas the absolute value of their 3-momentum is
determined from the energy deposited in the calorimeters.  
This energy deposition is required to geometrically match the track and 
the deposition's lateral shape must be consistent with that expected 
from electrons.
The additional transverse calorimeter energy deposited in a cone of $\Delta R = \sqrt{(\Delta\phi)^2+(\Delta\eta)^2}=0.4$
around each electron must be less than 10\% of its transverse energy.
The electrons are required to be separated by $\Delta R>0.4$ 
and to have the $z$ coordinate of their tracks at the origin within $|\Delta z|<5$ cm.
The average $z$ position of the two tracks must be within 4 cm of an interaction vertex.
Finally, each electron must have a track with an impact parameter less than 0.02 cm, if the track
is reconstructed including information from the silicon detector, or less than 0.2 cm otherwise. 
The forward electrons can be linked to special silicon-only tracks in order to extend the 
reach in $\eta$ beyond that of the drift chamber coverage.
Due to the limitation in determining the charge of forward electrons, no opposite-charge requirement is imposed.
After the above selection, we retain 240 224 dielectrons in the $Z$-boson resonance ($76<M_{ee}<106$~GeV/$c^2$)
and 801 dielectrons with $M_{ee}>200$~GeV/$c^2$. 

The main SM dielectron background to a high-mass dielectron signal is the Drell-Yan (DY) process $q\bar{q} \rightarrow Z/\gamma^* \rightarrow e^+ e^-$.  Secondary electroweak backgrounds come 
from diboson production ($WW$, $WZ$, $ZZ$, $W\gamma^*$) with subsequent electronic
decays.  The main hadronic background contributing to the dielectron candidate sample
is the production of $W$+jets, where the $W$ boson decays to an electron and the jet is misidentified
as an electron (``fake'' electron).  Finally, $t\bar{t}$ events decaying to dielectrons 
are also considered.

The DY, diboson, and $t\bar{t}$ backgrounds are estimated with Monte Carlo (MC) simulation,
using {\sc pythia} \cite{pythia}, running with the CTEQ5L \cite{cteq} parton distribution functions,
and the CDF {\sc geant}-based \cite{geant} detector simulator.  
The MC events are normalized on an event-by-event basis using the theoretical next-to-leading-order 
cross sections \cite{cross}, event trigger efficiencies, electron identification scale factors and 
data luminosity.  
Good agreement between SM expectation and CDF data in the $Z$ resonance
will validate our DY prediction.  The diboson
background estimation is validated in a trilepton $ee+e/\mu$ control region with $\met>15$ GeV \cite{footnote}, where the diboson contribution
is significant.  The $t\bar{t}$ dielectron background is validated in a control region with two or more hadronic jets, $H_T>200 $~GeV \cite{footnote2}, 
and $\met>20$~GeV, where the top-quark pair production is the dominant process.

The hadronic (QCD) background is estimated using CDF data, by selecting events 
with an identified electron and applying to every well-reconstructed and fiducial jet a probability to be misidentified  as an electron.  
The probability for a jet to fake an electron is determined from jet-rich CDF data \cite{marcelo}.
The QCD background estimation is validated in the intermediate-mass ($20<M_{ee}<76$~GeV/$c^2$) control region,
as well as in the trilepton ($76<M_{ee}<106$~GeV/$c^2$) and high-mass $t\bar{t}$ control regions described above.
 
The main sources of systematic uncertainty on the MC-estimated backgrounds \cite{syst}
are the theoretical cross sections (an 8\% effect on the event yields),
the luminosity (6\%), the lepton-ID efficiency (2\%), the parton distribution functions (2\%), and the trigger efficiency (0.5\%).
The total MC systematic uncertainty on the expected event yield is $\sim 10$ \%.  The 
respective QCD-background systematic uncertainty is $\sim 50$ \%, which comes from the variation
in the measurement of the fake probabilities using different jet-rich CDF datasets
triggered with varied jet-energy thresholds.  
Since the SM background is very low in the high dielectron-mass region of interest, the results are not overly sensitive to the systematic uncertainties.

Figure 1(a) shows the $M_{ee}$ spectrum for SM expectation and CDF data.  We
observe good agreement in the $Z$-boson resonance region, which validates our efficiencies,
scale factors, data luminosity and theoretical cross sections.  We observe a long mass tail coming mainly from the DY process.  Figure 1(b) shows the $M_{ee}$ spectrum between 200 and 600~GeV/$c^2$; above 200~GeV/$c^2$, we expect $793 \pm 96$ events and observe 801. The previously observed excess in the 225-250 GeV/$c^2$ region \cite{cdf_ee} has been reduced to a $<1$-$\sigma$ effect, in this larger dataset.
We observe an exceptional $M_{ee}=960$~GeV/$c^2$ candidate event, the highest-mass dielectron event
ever observed.  Calorimeter resolution induces an uncertainty on the dielectron mass of 16~GeV/$c^2$ for this event.  
The two electrons have transverse momenta of 482 and 468~GeV/$c$;
they originate from the same primary vertex and are oppositely charged.  The event is characterized
by very low hadronic activity (no jets are reconstructed) and by a low missing transverse energy 
of 17~GeV -- separated by 23 degrees in $\phi$ from the 468~GeV/$c$ electron -- coming most probably from the resolution
of the calorimeter.
\begin{figure}[!]
\hspace{-3.5cm}
\begin{minipage}[!]{2.in}
\begin{center}
\includegraphics[scale=0.44]{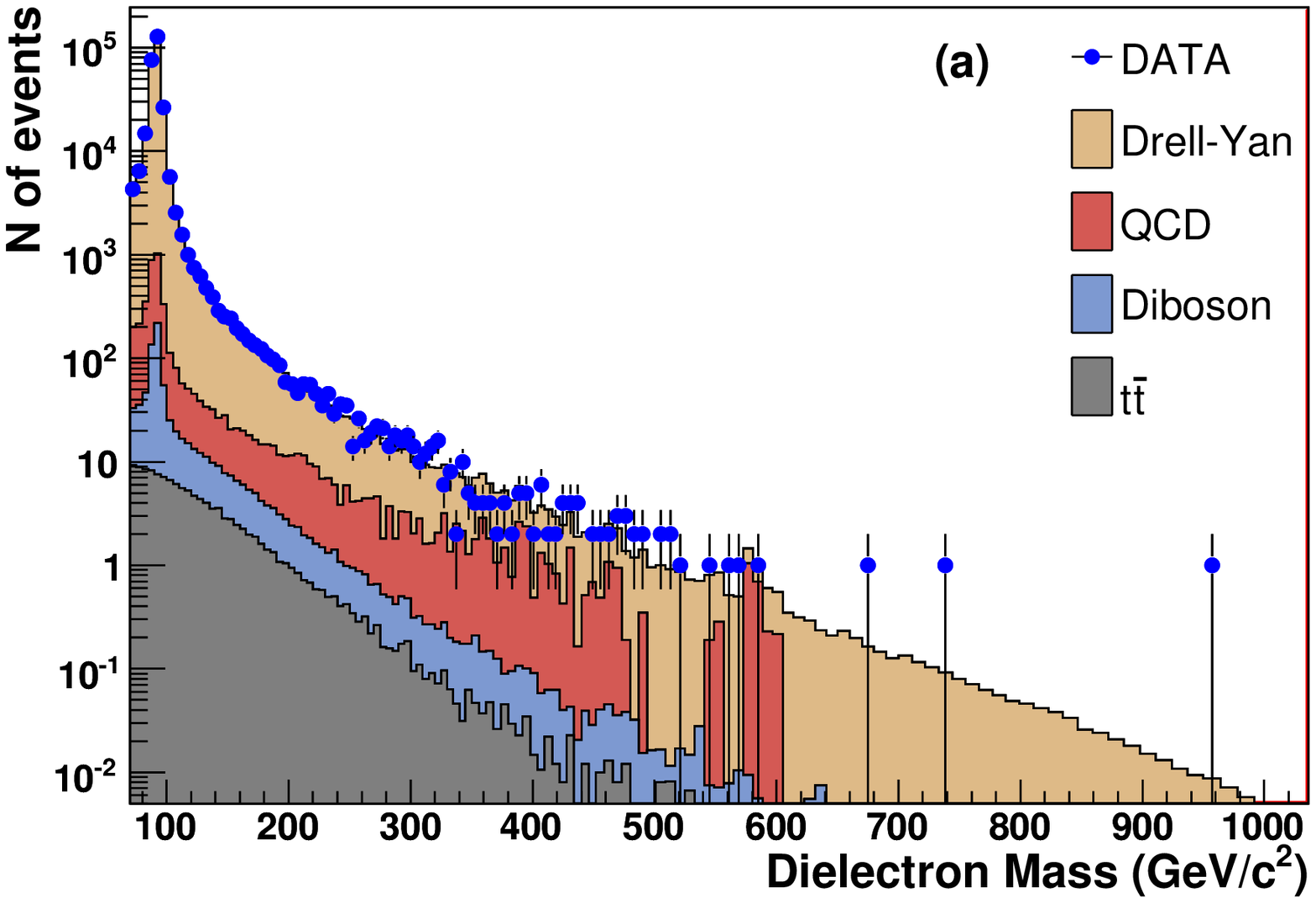}
\end{center}
\end{minipage}
\\
\hspace{-3.5cm}
\begin{minipage}[!]{2.in}
\begin{center}
\includegraphics[scale=0.44]{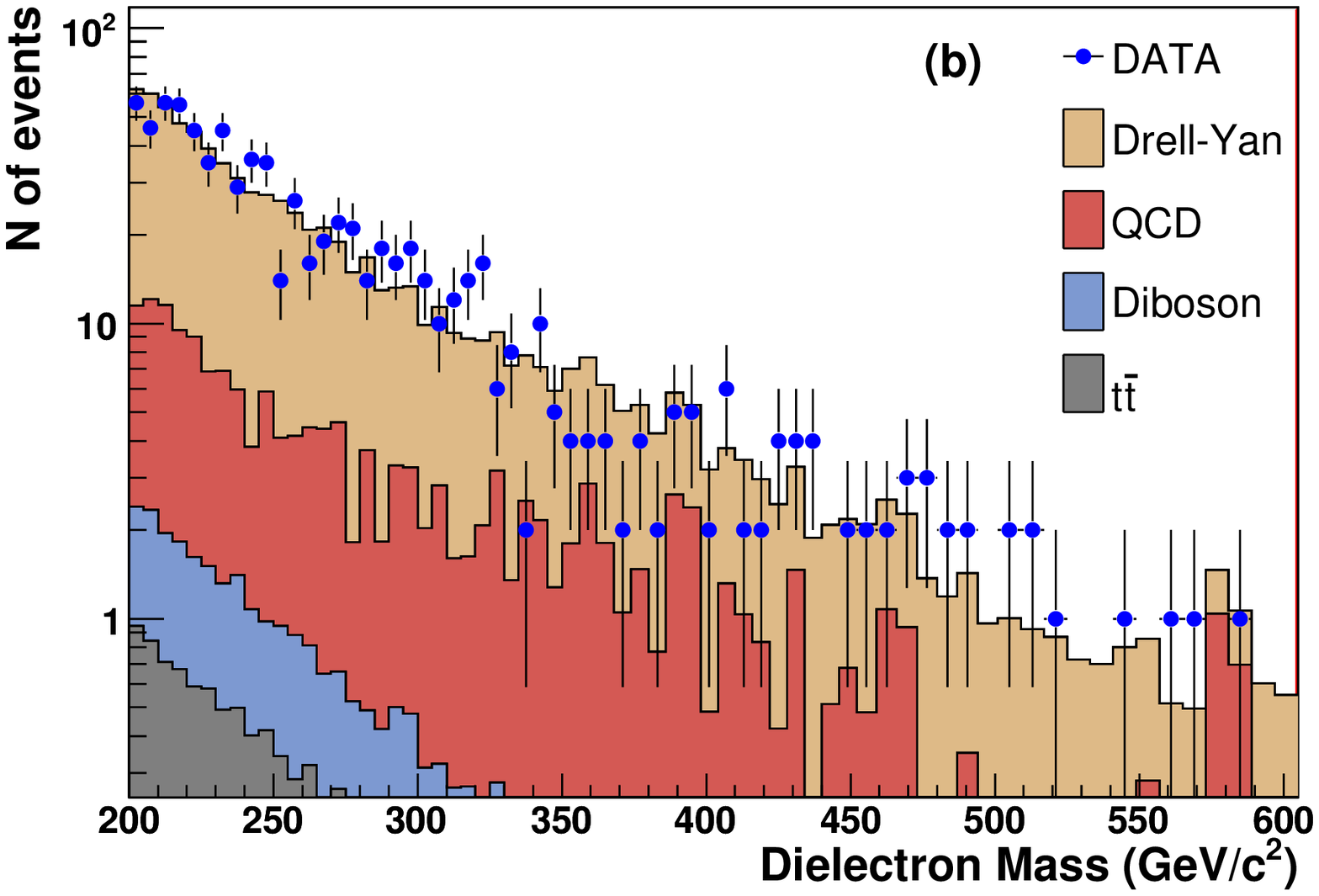}
\end{center}
\end{minipage}
\caption{(a) The dielectron mass distribution for the SM background (stacked histograms of DY, QCD, diboson, and $t\bar{t}$)
and the CDF data.  The QCD background is derived from CDF data.  The electroweak backgrounds are estimated
using MC normalized to the data luminosity times theoretical cross sections, without fitting to any part of the dielectron data distribution.  (b) The dielectron mass distribution between 200 and 600~GeV/$c^2$.  
}
\label{mass}
\end{figure}
\begin{figure}[!]
\begin{center}
\includegraphics[scale=0.44]{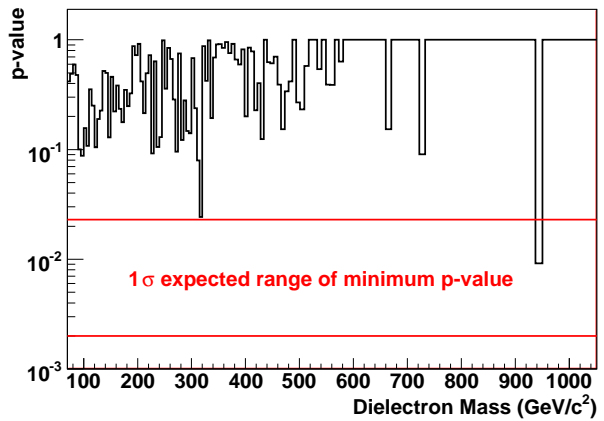}
\end{center}
\caption{The $p$-value as a function of the dielectron mass.  The 1-$\sigma$ range of expected minimum $p$-value
seen at any bin is shown.} 
\label{pval}
\end{figure}

\begin{figure}[!]
\hspace{-4cm}
\begin{minipage}[!]{2in}
\begin{center}
\includegraphics[scale=0.45]{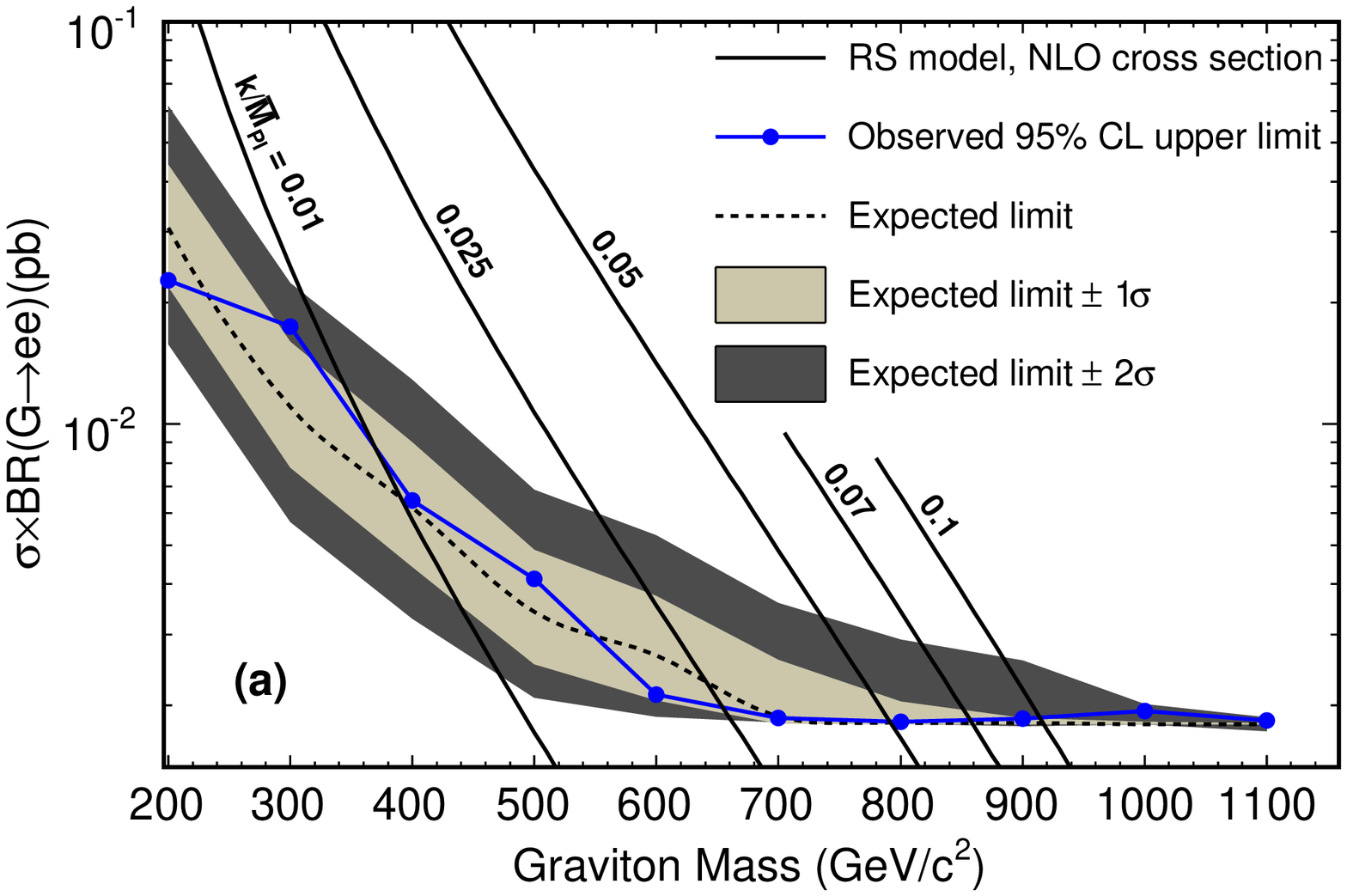}
\end{center}
\end{minipage}
\\
\hspace{-4cm}
\begin{minipage}[!]{2in}
\begin{center}
\includegraphics[scale=0.45]{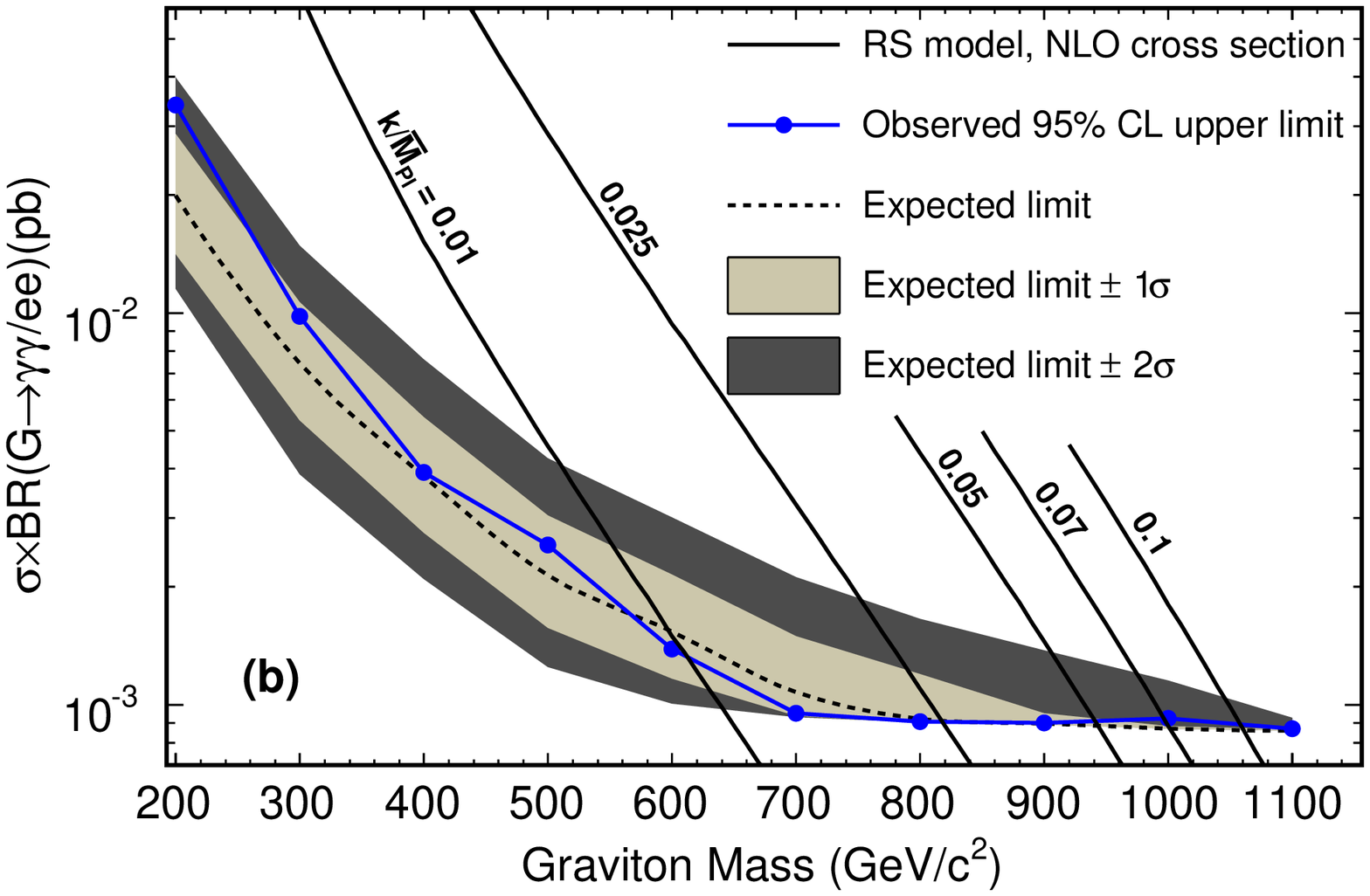}
\end{center}
\end{minipage}
\caption{(a) The 95\% $\cl$ upper limit on the lightest RS-graviton production 
cross section from the dielectron analysis, as a function of the graviton's mass.  Theoretical curves are shown for five values of the coupling $k/\overline{M}_{Pl}$ from 0.01 to 0.1.  The intersections give the RS-graviton lower mass limits for these five hypotheses. (b) The respective cross section and mass limits for the combination of the dielectron and diphoton analyses.  The inclusion of the dielectron analysis improves the $k/\overline{M}_{Pl}=0.1$ limit by $\sim$95~GeV/$c^2$.} 
\label{limits_1d}
\end{figure}
In order to assess the agreement between SM and observation, for every mass bin $i$
we define a $p$-value as the probability to observe at least $N_{data}^{i}$ events, 
when $N_{SM}^{i}$ events are expected.  We determine this probability by generating pseudoexperiments
and counting the number of times when $N_{pseudo}^{i} \ge N_{data}^{i}$.  Here $N_{pseudo}^{i}$ follows a Poisson distribution 
convoluted with a Gaussian of mean equal to $N_{SM}^{i}$ and standard deviation equal to the SM systematic 
uncertainty for the $i$-th bin. Figure \ref{pval} shows the $p$-value as a function of the $M_{ee}$.
The band defines the 1-$\sigma$ expected range of the minimum $p$-value that could be
observed in any mass bin.  This range is calculated from the probability distribution of minimum $p$-values
determined with the use of pseudoexperiments. 
The probability to observe a SM event with $M_{ee}\ge 960$~GeV/$c^2$ is 4\%.
Above 600~GeV/$c^2$, we expect $3.6 \pm 0.4$ events and observe three. 
The second most significant excess of events is observed at $\sim$320~GeV/$c^2$, as shown in Fig.\ \ref{pval}.

The measured $M_{ee}$ spectrum can be used to set a limit on RS-graviton production.
Here we parametrize the RS model using the mass of the lightest RS graviton 
($m_1$) and the dimensionless parameter $\sqrt{8\pi} k/M_{Pl} \equiv k/\overline{M}_{Pl}$,
where $k$ is the curvature scale of the extra dimension and $M_{Pl}$ is the Planck mass. 
RS-graviton signal MC is generated using {\sc pythia} and values of $m_1$ from 200 to 1100~GeV/$c^2$.   
The signal-MC events are normalized in the same manner as the background-MC events.
The leading-order {\sc pythia} cross section is multiplied by a scale (``$K$-factor'') 
to correct for next-to-leading-order effects.
The CDF acceptance for the dielectron RS-graviton 
signal, determined using MC, is $\sim$27\%.  The acceptance is approximately independent of the RS-graviton 
mass due to the inclusion of forward electrons, as lower-mass gravitons ($m_1<700$~GeV/$c^2$)
would decay to at least one forward lepton more often than higher-mass gravitons. 
The RS-graviton decays to two jets (branching ratio BR=70\%), two charged leptons (BR=6\%),
two neutrinos  (BR=6\%), two photons (BR=4\%) and two weak bosons (BR=14\%).
For graviton masses above 200 GeV/$c^2$, these branching ratios are not graviton-mass-dependent.  
Given the considerable dijet background in a hadron-collider environment, and
the low leptonic branching fraction of the weak bosons, the prompt dilepton and diphoton final states 
offer the greatest sensitivity.  The inclusion of dielectrons
in the diphoton RS search \cite{cdf_gg} results in a 50\% increase in the rate of potentially produced signal.  
Here we present the cross section and mass limits for the dielectron channel alone and for the combined
dielectron+diphoton channel.

\begin{figure}[!]
\begin{center}
\hspace{-0.5cm}
\includegraphics[scale=0.44]{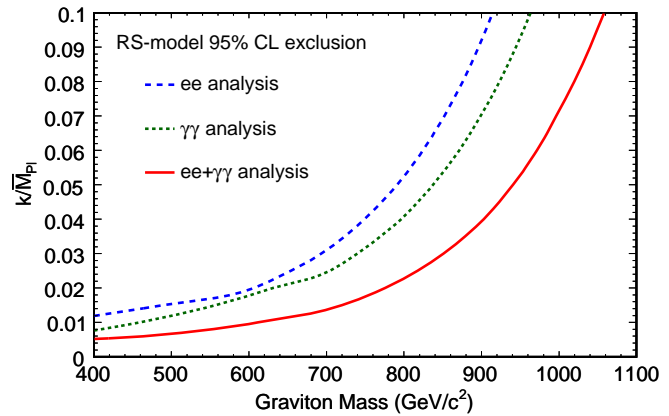}
\end{center}
\caption{The area excluded in the $k/\overline{M}_{Pl}$ vs.\ $m_1$ RS-model parameter space by the 5.7 fb$^{-1}$ dielectron analysis (area above long-dashed line), the 5.4 fb$^{-1}$ diphoton analysis (area above short-dashed line), and their combination (area above solid line).}
\label{limit_2d}
\end{figure}
Figure 3(a) shows the 95\% confidence level ($\cl$) cross-section ($\sigma \times {\rm BR}(G\rightarrow ee)$) 
upper limit as a function of the lightest RS-graviton mass $m_1$ along with five theoretical cross-section 
curves for $k/\overline{M}_{Pl}=0.01$ to 0.1, a theoretically interesting range that would provide a solution for the fine-tuning problem.  
The limits are set using a frequentist method that
compares the background-only with the signal+background hypotheses, taking into account correlated background systematic uncertainties \cite{tomjunk,cdf_gg}.  
The intersection of the cross-section exclusion limit 
with the theoretical cross-section curves gives the 95\% $\cl$ lower limit on $m_1$ for the respective coupling.
For $k/\overline{M}_{Pl}=0.1$, the dielectron-only $m_1$ lower limit is 914~GeV/$c^2$, 
if we use proper mass-dependent RS-graviton $K$-factors \cite{mathews,kumar}, and 935~GeV/$c^2$ assuming a 
fixed $K$-factor of 1.54 \cite{d0}, for comparison with previous results \cite{d0,cdf_ee}.  For $m_1>700$~GeV/$c^2$, the dielectron-only analysis
excludes cross sections greater than 2 fb at the 95\% $\cll$.  Above 700~GeV/$c^2$ the expected background number of events are consistent with zero, leading to a graviton-mass-independent expected limit in the investigated range of graviton masses.  
Figure 3(b) shows the 95\% $\cl$ cross-section ($\sigma \times {\rm BR}(G\rightarrow ee/\gamma\gamma)$) upper limit from the combination of the dielectron and diphoton analyses.  The two analyses are combined as independent channels bound by a common production cross section and some correlated systematic uncertainties \cite{tomjunk}.
For $k/\overline{M}_{Pl}=0.1$, the combined $m_1$ lower limit is 1058~GeV/$c^2$, if we use proper mass-dependent 
RS-graviton $K$-factors, and 1092~GeV/$c^2$ assuming a fixed $K$-factor of 1.54. 
For $m_1>700$~GeV/$c^2$, the combined $ee+\gamma\gamma$ analysis excludes cross sections greater than 0.9 fb at the 95\% $\cll$.

Figure \ref{limit_2d} shows the exclusions in the $k/\overline{M}_{Pl}$ vs. $m_1$ parameter space for the dielectron 
analysis, the diphoton analysis and their combination.
For $k/\overline{M}_{Pl}=0.1$, the inclusion of electrons improves the previous CDF RS-graviton limit \cite{cdf_gg} by $\sim$95~GeV/$c^2$
and surpasses the previous published \cite{d0} best limit by $\sim$40~GeV/$c^2$.  Table \ref{table2} summarizes the combined $ee+\gamma\gamma$ RS-graviton lower mass limits for different values of $k/\overline{M}_{Pl}$.
\begin{table}[!]
\begin{center}
\caption{The 95\% $\cl$ RS-graviton lower mass limits from the combined $ee+\gamma\gamma$ analysis, for five values of $k/\overline{M}_{Pl}$.}
\begin{tabular}{cc}
\hline \hline
$k/\overline{M}_{Pl}$ & Mass limit (GeV$/c^2$)\\
\hline\hline
0.01 & 612\\
0.025& 818\\
0.05& 941\\
0.07& 997\\
0.1& 1058\\
\hline \hline
\end{tabular}
\label{table2}
\end{center}
\end{table}

In summary, we present a search for high-mass dielectron resonances using CDF data corresponding to an integrated luminosity of 5.7 fb$^{-1}$.
We do not observe any significant discrepancies from the expected SM prediction, while we detect 
the highest-mass dielectron event ever observed, at 960~GeV/$c^2$.  We use the dielectron mass spectrum
to set 95\% $\cl$ limits on the RS-graviton production, for several values of the $k/\overline{M}_{Pl}$ parameter.
The graviton lower mass limit from the combined dielectron and diphoton analyses and for $k/\overline{M}_{Pl}=0.1$
is 1058 or 1092~GeV/$c^2$, depending on the assumption on the NLO cross sections, making this result the most stringent to date.

We thank M.\ C.\ Kumar, P.\ Mathews, and V.\ Ravindran for the calculation of the mass-dependent RS-graviton $K$-factors.
We also thank the Fermilab staff and the technical staffs of the participating institutions for their vital contributions. This work was supported by the U.S. Department of Energy and National Science Foundation; the Italian Istituto Nazionale di Fisica Nucleare; the Ministry of Education, Culture, Sports, Science and Technology of Japan; the Natural Sciences and Engineering Research Council of Canada; the National Science Council of the Republic of China; the Swiss National Science Foundation; the A.P. Sloan Foundation; the Bundesministerium f\"ur Bildung und Forschung, Germany; the Korean World Class University Program, the National Research Foundation of Korea; the Science and Technology Facilities Council and the Royal Society, UK; the Institut National de Physique Nucleaire et Physique des Particules/CNRS; the Russian Foundation for Basic Research; the Ministerio de Ciencia e Innovaci\'{o}n, and Programa Consolider-Ingenio 2010, Spain; the Slovak R\&D Agency; the Academy of Finland; and the Australian Research Council (ARC).


\begin{thebibliography}{39}
\bibitem[{foo({\natexlab{a}})}]{footnote}
\bibinfo{note}{The missing transverse energy vector $\vec{\met}$ is defined as 
$-(\sum_i \vec{E}_{i})_T$, where $\vec{E}_{i}$ has magnitude equal to the energy deposited in the 
$i{\rm th}$ calorimeter tower and direction perpendicular to the beam axis and pointing to that calorimeter 
tower.  The $\met$ is corrected for the presence of muons, because they do not deposit their energy in the
calorimeters.}
\bibitem[{foo({\natexlab{a}})}]{footnote2}
\bibinfo{note}{$H_T$ is defined as the scalar sum of the transverse momenta 
of jets, leptons and $\met$}.
\end{thebibliography}
\end{document}